\documentclass[twoside,leqno,twocolumn]{article}
\usepackage{ltexpprt}

\usepackage{color}
\usepackage{amsmath}
\usepackage{graphicx}
\usepackage{subfigure}

\begin{document}



\title{\Large Network-based Anomaly Detection for Insider Trading
\thanks{Work in part funded by George Mason University, Vice President for Research Office,  Multidisciplinary Research Award.}}
\author{Adarsh Kulkarni\thanks{George Mason University.} \\
\and
Priya Mani\thanks{George Mason University.} \\
\and
\and Carlotta Domeniconi\thanks{George Mason University.}}
\date{}

\maketitle


\begin{abstract} \small\baselineskip=9pt 
Insider trading is one of the numerous white
collar crimes that can contribute to the instability of the economy. Traditionally, the detection of illegal insider trades has been a human-driven process. In this paper, we collect the insider trade filings made available by 
the US Securities and Exchange Commission’s (SEC) through the EDGAR system, with the aim of initiating an automated large-scale and data-driven approach to the problem of identifying illegal insider tradings. 

The goal of the study is the identification of interesting patterns, which can be indicators of potential anomalies. We use the collected data to construct networks that capture the relationship between trading behaviors of insiders. We explore different ways of
building networks from insider trading data, and argue for a need of a structure that is capable of capturing higher order relationships among traders. Our results suggest the discovery of interesting patterns. 

\end{abstract}

\section{Introduction}

Financial markets are notoriously difficult to understand, which makes the tracking of white collar crimes such as Illegal Insider Trading very challenging. As such, the detection of illegal insider trades has been a characteristically human-driven process. We aim to challenge this trend by taking a big data approach to the problem and investigate graph-based data mining techniques to identify patterns of illegal insider trades. 
Through the US Securities and Exchange Commission's (SEC) Electronic Data Gathering, Analysis and Retrieval system (EDGAR), insider trade filings have been made public, providing a substantial source of data for research.  

	Insider trading is a subset of the numerous white collar crimes that can contribute to the instability of the economy. Financial crimes and mishandling of billions of dollars are key issues in our society that have been blamed for the financial crises we have seen over the last decade. Working to detect and curb financial crimes like illegal insider trading is a clear interest for both the government and the people. 
	
	Insider Trading, by formal definition, is not always illegal. Generally, insiders in a company tend to be either Officers (CEO, CFO), large shareholders ($>$10\%) or members of the Board of Directors. For these people, a substantial percentage of their compensation comes from stock and option awards. When they feel the need to liquidate their holdings, they file with the SEC and dispose of their shares. This process becomes illegal when the insider leverages information that only he or she may possess in order to trade stock at an unfair profit. For example, if the CEO of a company knows the stock will rise after announcing that they have surpassed their quarterly goals, and decides to buy stock before that announcement, the CEO is misusing his or her information to gain an unfair advantage.	Currently, the SEC requires all insiders to file a Form 4 whenever they acquire or dispose of their company's stock. These forms require the insider to declare how much they are trading, what price they are trading at, and how large their remaining holdings after the trade will be.  These are the filings that the SEC publishes through the EDGAR system. 

In this work, we leverage this data to construct graphs that capture the relationship between  trading behaviors of insiders. The study aims at identifying interesting patterns, which can be indicators of potential anomalies. We first apply the technique introduced in \cite{SNAM14} to our data. We then explore additional ways of building networks from insider trading data, and argue for a need of a structure that is capable of capturing higher order relationships among traders. While preliminary, our results suggest the discovery of interesting patterns. We also identify future challenges to be addressed and possible directions to tackle them.

\section{Related Work}
Some attempts to automate the detection of illegal insider trading have been made. Goldberg et al. \cite{Goldberg03} have found that over 85\% of Insider Trading cases are correlated to five different types of news: Product announcements, Earnings announcements, Regulatory approval or denials, Mergers and acquisitions or Research reports, which they collectively refer to as PERM-R events. The task of maintaining surveillance over trading activity is gargantuan, with ~5.5 million trades being found to be of interest by the Insider Trading and Fraud Teams at NASD, which is an organization that is self-run and keeps watch over multiple security markets, the most notable being the Nasdaq Stock Exchange \cite{Goldberg03}. SONAR was designed by Goldberg et al. to automate as much of the manual process that NASD goes through every day. Consolidating data from numerous data sources such as Reuters, Bloomberg and the Dow Jones, it uses Natural Language Processing to analyze 8,000-10,000 articles and correlate them with EDGAR filings as well as the rest of the market activity. Based 
on all of this data, SONAR tries to flag suspicious trades that it believes must be more thoroughly investigated. 

Another study focuses on the detection of illegal insider trading in the options market \cite{Donoho}. This study investigated the patterns created by insider trading in the option markets. They focused on analyzing options as they are much less frequently traded, compared to stocks. 
They conducted multiple case studies and found that whenever there was a PERM-R announcement similar to those described in \cite{Goldberg03}, option trading volume would spike. One of their case studies described the acquisition of Pharmacia by Pfizer. Pharmacia stock opened 20\% higher than its last closing price after announcing the acquisition over the weekend. Till then call volume had been steadily below 1,500 trades, but in the days leading up to the announcement it rose to above 8,000. 
Options bought for \$0.55 before the announcement could have been sold 3 days later for \$4.10, a 650\% increase. While Donoho concentrates on the option markets which is indeed ripe for analysis, in this work we focus on the standard stock market, with the intent of sheding some light on potential patterns leading to illegal insider trades. 

This paper was inspired by the recent work of Tamersoy et al. \cite{SNAM14}. The authors explored the relationship between trading behavior and insiders' roles, their companies' sectors, and their relationships to other insiders. They also performed anomaly detection over a network they created over all insiders in their dataset. They found that many insiders are part of cliques, where all the trading behaviors of the members of the clique are very similar, as well as that insiders tend to have an abnormally high profitability of trades. Our findings concur with these observations, but we also aim to overcome some of the limitations of the approach in \cite{SNAM14}. In particular, in this work we explore additional ways of building networks from insider trading data, and argue for a need of a structure that is capable of capturing higher order relationships, e.g. hyper-graphs.

A variety of methods has been introduced for graph-based anomaly detection, although very little has been done in the area of illegal insider trading. The approach depends on the nature of the graph, e.g. attributed vs. non-attributed, or static vs. dynamic. A complete overview of such methods is beyond the scope of this paper. A survey of the various approaches is found in \cite{Akoglu-survey}.

\section{Data}

There are numerous online resources and websites that follow the EDGAR RSS Feed, and mine it for its data. We chose to use  Insider Monkey\footnote{http://www.insidermonkey.com/insider-trading/sales/}, which monitors EDGAR in real time and parses Form 4 filings as they are processed by the SEC and made public. From Insider Monkey, we scraped ~1.1M insider trades, as well as all insider positions. For example, if we consider the case of Salesforce CEO Marc Benioff, Insider Monkey provides us with his entire record of stock trades for Salesforce (CRM), as well as the fact that he is Chairman and CEO of Salesforce, a Large Shareholder at Fitbit (FIT) and a member of the Board of Directors at Cisco Systems (CSCO). 

The data was  organized as a MySQL database and included the names of traders, their positions, and the companies whose stocks they traded. The historical price of the stocks were also scraped from Google Finance. For every ticker in Nasdaq and NYSE, we have pricing data going all the way back to the Initial Public Offering of the company, which spans as far back as the 1980s for older corporations.
The summary statistics of the data is shown in Table~\ref{stat}.


{
\begin{table}[h]	
\centering
	\begin{tabular}[l]{l|l}
	Insiders & 70,408 \\
	Companies & 12,485 \\
	Sale Transactions & 757,194\\
	Purchase Transactions & 311,013 \\
	\end{tabular}
	\caption{Global Statistics}
\label{stat}		
\end{table}
}

%

\section{Network-based Anomaly Detection}

\subsection{Building Networks of Insiders} 
\begin{figure}[h]
\centering
		\includegraphics[scale=0.3]{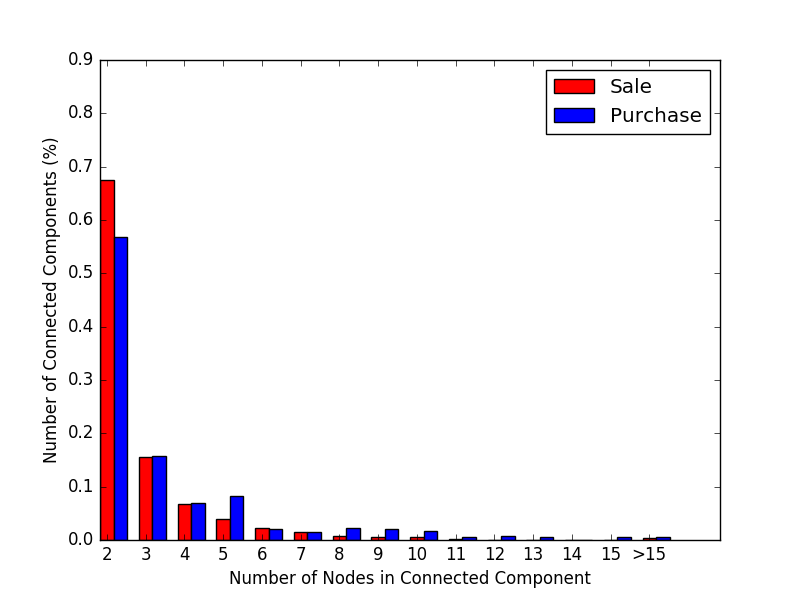}
		\caption{Distribution of Connected Components (based on $S$)}
		\label{hist_conn_comp}
		\end{figure}

As a preliminary analysis we followed an approach similar to \cite{SNAM14} where networks are constructed based on the trading behaviors of insiders and analyzed for anomalies among connected components. We constructed purchase and sale networks with insiders as nodes. Only insiders with at least 5 trades were considered. Edges were added based on similarity scores as defined below. Initially, we used a similarity function that takes into account the dates on which the insiders traded, and what proportion of those dates were common among the two insiders, as done in \cite{SNAM14}. 
Specifically, let $X_C$ and $Y_C$ be two traders of company $C$, each represented as the set of dates on which he traded. Their similarity is computed as follows:
$$S(X_{C},Y_{C}) = \frac{(\sum\limits_{i=1}^{|X_{c}|} \sum\limits_{j=1}^{|Y_{C}|}\text{I}(x_{i},y_{j}))^2}{|X_{C}| \times |Y_{C}|}$$
where $I()=1$ when the arguments have the same value, and $0$ otherwise.
Table~\ref{net_stat} and Figure~\ref{hist_conn_comp} show the statistics for the purchase and sale networks, corresponding to a threshold value of $0.5$ on the similarity $S$. The statistics include all connected components of at least size 2 (isolated nodes were discarded).
{
\begin{table}[h]	
\centering
	
			\begin{tabular}{rrrr}
		 	\hline
		 	{\it Network} & {\it Nodes} & {\it Edges} & {\it Connected Components}\\
			\hline
			Sale & 1,508 & 1,943 & 543 \\
			Purchase & 1,414 & 3,263 & 401 \\
			\hline
			\end{tabular}
		\caption{Network Statistics (based on $S$)}	
\label{net_stat}		
\end{table}
}

\begin{figure}[h]
\centering
		\includegraphics[scale=0.3]{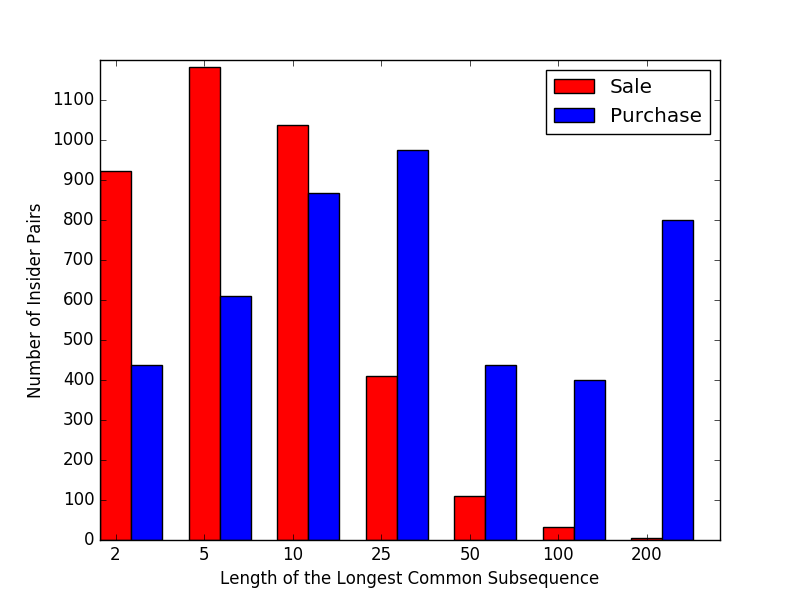}
		\caption{Distribution of Longest Common Sub-sequences.}
		\label{lcs_dist}
		\end{figure}

\begin{figure}[h]
\centering
		\includegraphics[scale=0.4]{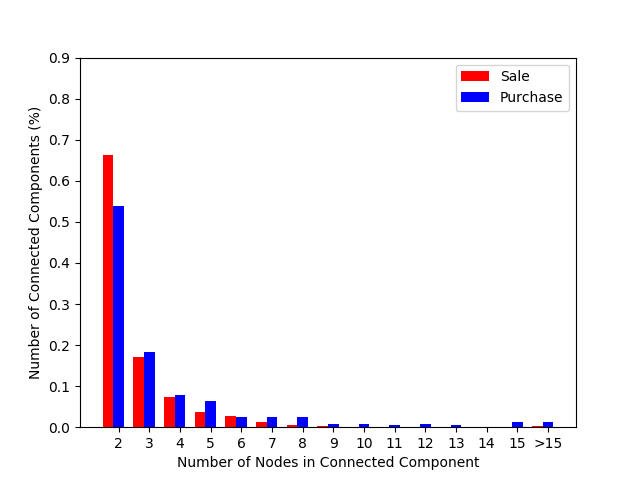}
		\caption{Distribution of Connected Components ({\textit{LCS-based}})}
		\label{lcs_hist_conn_comp}
		\end{figure}

We observe that the similarity $S$ above does not account for the temporal ordering of trading dates. As such, two traders are considered equally similar whether they share 7 consecutive trading dates, or 7 dates sparsely spread out in time. Intuitively, though, we may want to consider the first case as a stronger indication of similarity. To account for temporal ordering, we constructed new sale and purchase networks, where a node is represented as the {\textit{sequence}} of his trading dates. If two traders (nodes) shared a sub-sequence of length at least $t$ (threshold), we added an edge between the two nodes. We call this construction {\textit{LCS-based}}. The threshold $t$ was chosen based on the distribution of the length of the longest common sub-sequences among traders in the sale and purchase networks (shown in Figure \ref{lcs_dist}). As a result, we set $t=5$ (corresponding to 75.09\% of insider pairs) and $t=10$ (corresponding to 71.8\% of insider pairs) for the sale and purchase networks respectively. Table~\ref{lcs_net_stat} and Figure~\ref{lcs_hist_conn_comp} show the statistics for the \textit{LCS-based} purchase and sale networks. A similar trend as before is observed.

{

\begin{table}[h]	
\centering
	
			\begin{tabular}{rrrr}
		 	\hline
		 	{\it Network} & {\it Nodes} & {\it Edges} & {\it Connected Components}\\
			\hline
			Sale & 1,885 & 2,178 & 689 \\
			Purchase & 886 & 2,701 & 239 \\
			\hline
			\end{tabular}
		\caption{Network Statistics ({\textit{LCS-based}})}	
\label{lcs_net_stat}		
\end{table}
}

\subsection{Approach}		
The egonets in the purchase and sale networks are  analysed for anomalies \cite{SNAM14,Akoglu2010}. Let $V_u$ be the number of nodes and $E_u$ the number of edges of the egonet corresponding to the ego node $u$. The plot of $V_u$ against $E_u$ across all egonets revealed a power law relationship. A least squares fit on the median values of $E_u$ was computed and outlier scores were assigned to each ego node. 
The outlier score measures the deviation of the ego node $u$ from the power law relationship, and is defined as \cite{SNAM14,Akoglu2010}:\\
$$\text{Score}(u) = \frac{\max(E_u,f(V_u))}{\min(E_u,f(V_u))} \times (\log(|E_u -f(V_u)|+1))$$
where $f(V_u)$ is the least squares fit on the median values of $E_u$.

A local outlier factor measuring the density of $u$ with respect to the density of its neighbors (as in LOF \cite{LOF}) was added to Score($u$) to obtain the Total Outlier Score, as done in \cite{Akoglu2010}:
$$\text{TotalOutlierScore}(u)  = \text{Score}(u) + \text{LOF}(u)$$

\subsection{Results}
\begin{figure}[h]
\begin{center}
			
\subfigure[] {  \includegraphics[width=35mm,height=30mm]{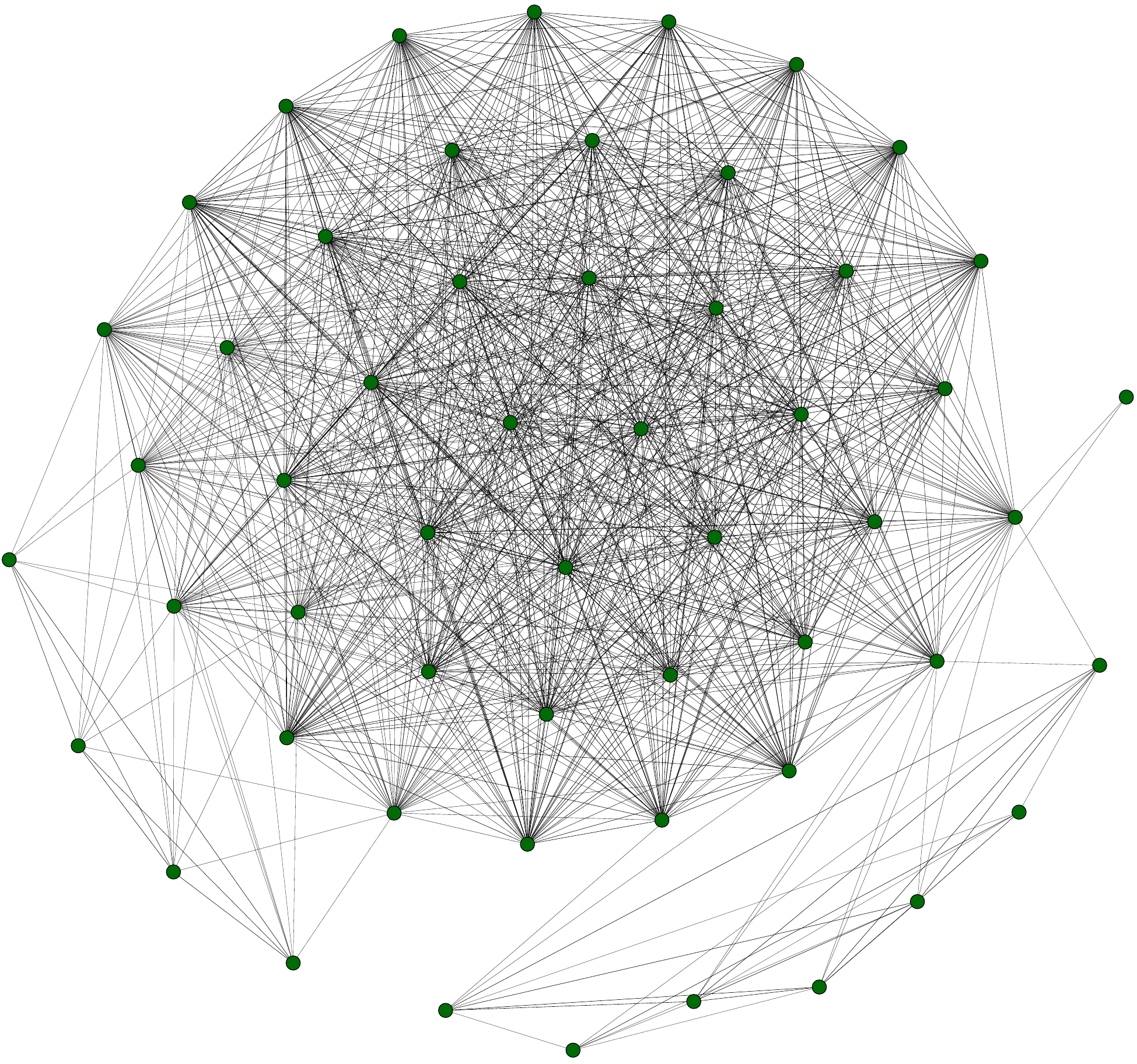}}
\subfigure[] {  \includegraphics[width=35mm,height=30mm]{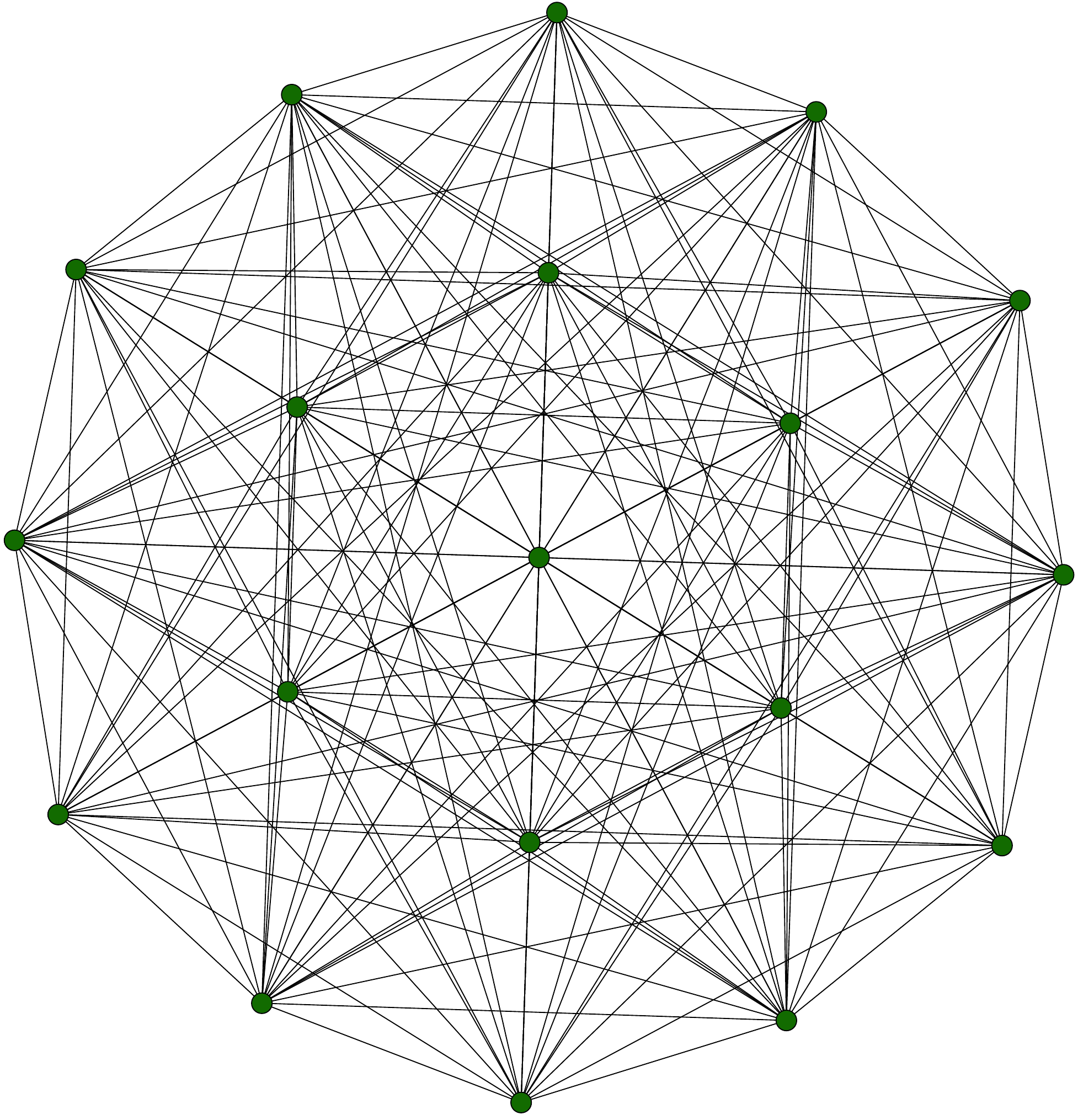}}
\end{center}
\caption{Connected Components: (a) Purchase: International Speedway Corporation; (b) Sale: Vantiv Inc.}
\label{conn_comp}
\end{figure}

Examples of connected components observed from the networks constructed using the similarity measure $S$ are depicted in Figure~\ref{conn_comp}. The figure shows that they are highly connected components, which is an indication of frequent pairwise similarities. 

\begin{figure}[h]
\begin{center}
\subfigure[] {  \includegraphics[width=70mm,height=55mm]{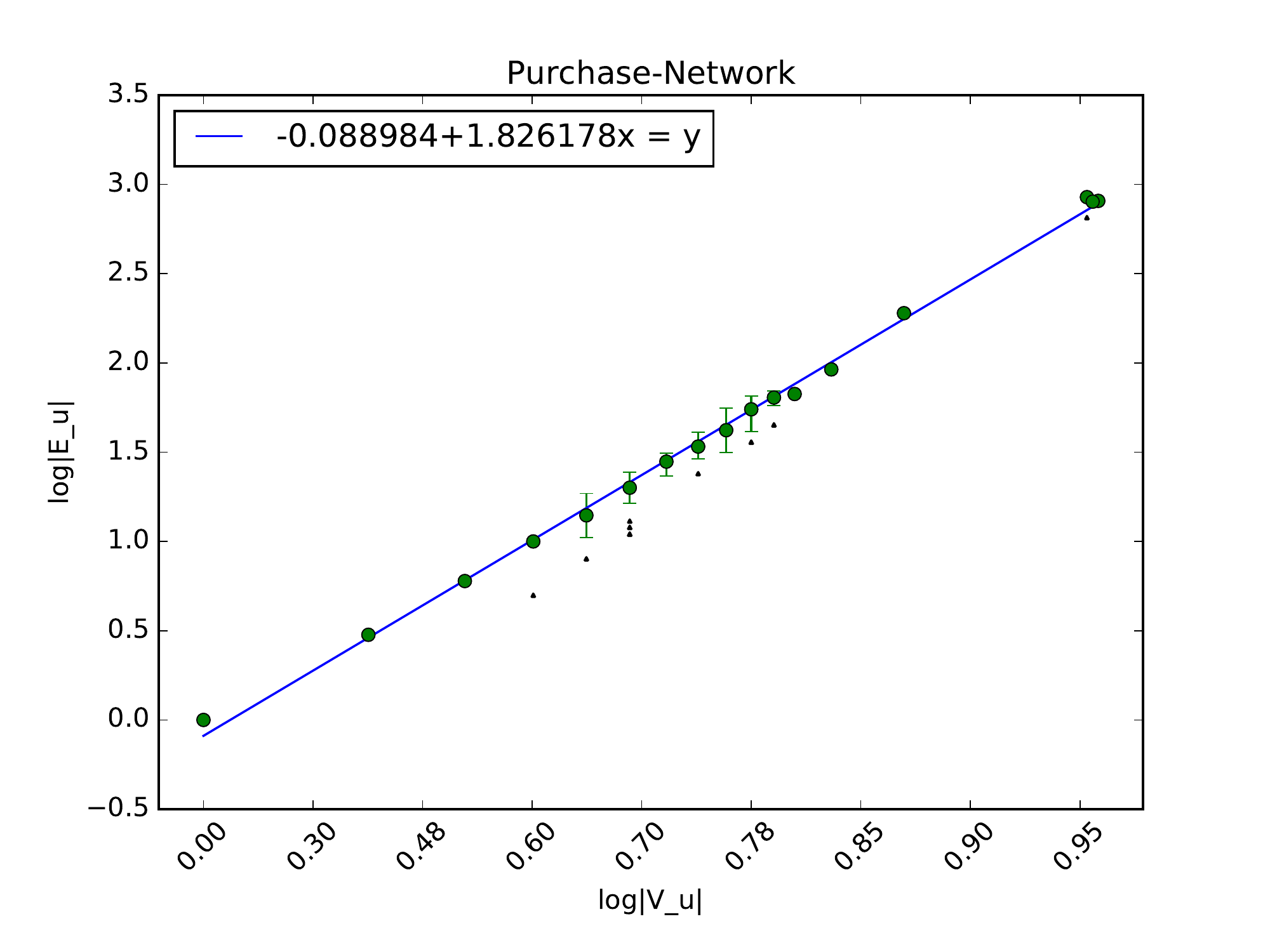}}
\subfigure[] {  \includegraphics[width=70mm,height=55mm]{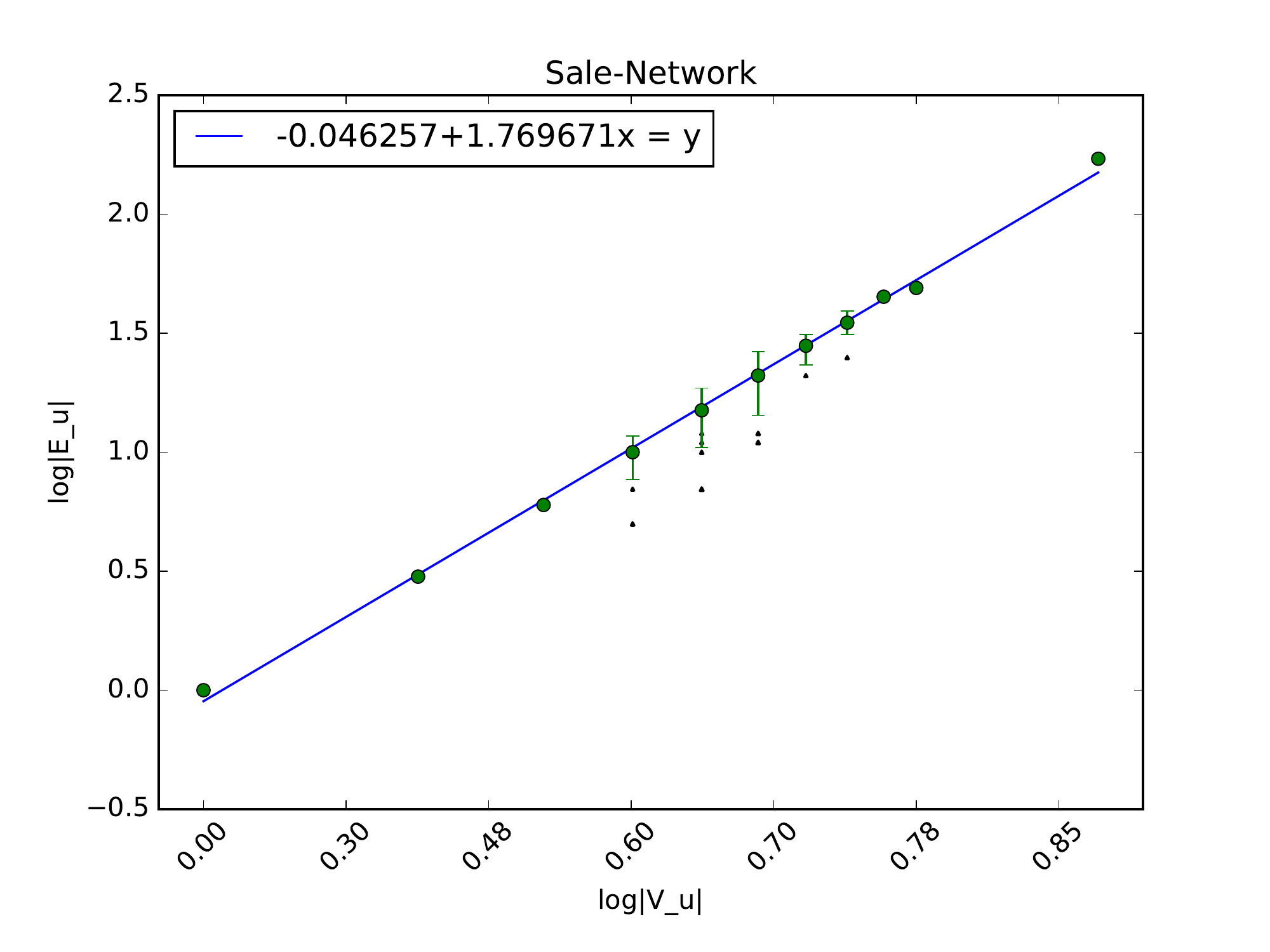}}
\end{center}
\caption{Power Law Fitting and Anomaly Detection: (a) Purchase; (b)  Sale. }
\label{lsfit}
\end{figure}

\begin{figure}
\begin{center}
 \subfigure[] {\includegraphics[width=65mm,height=50mm]{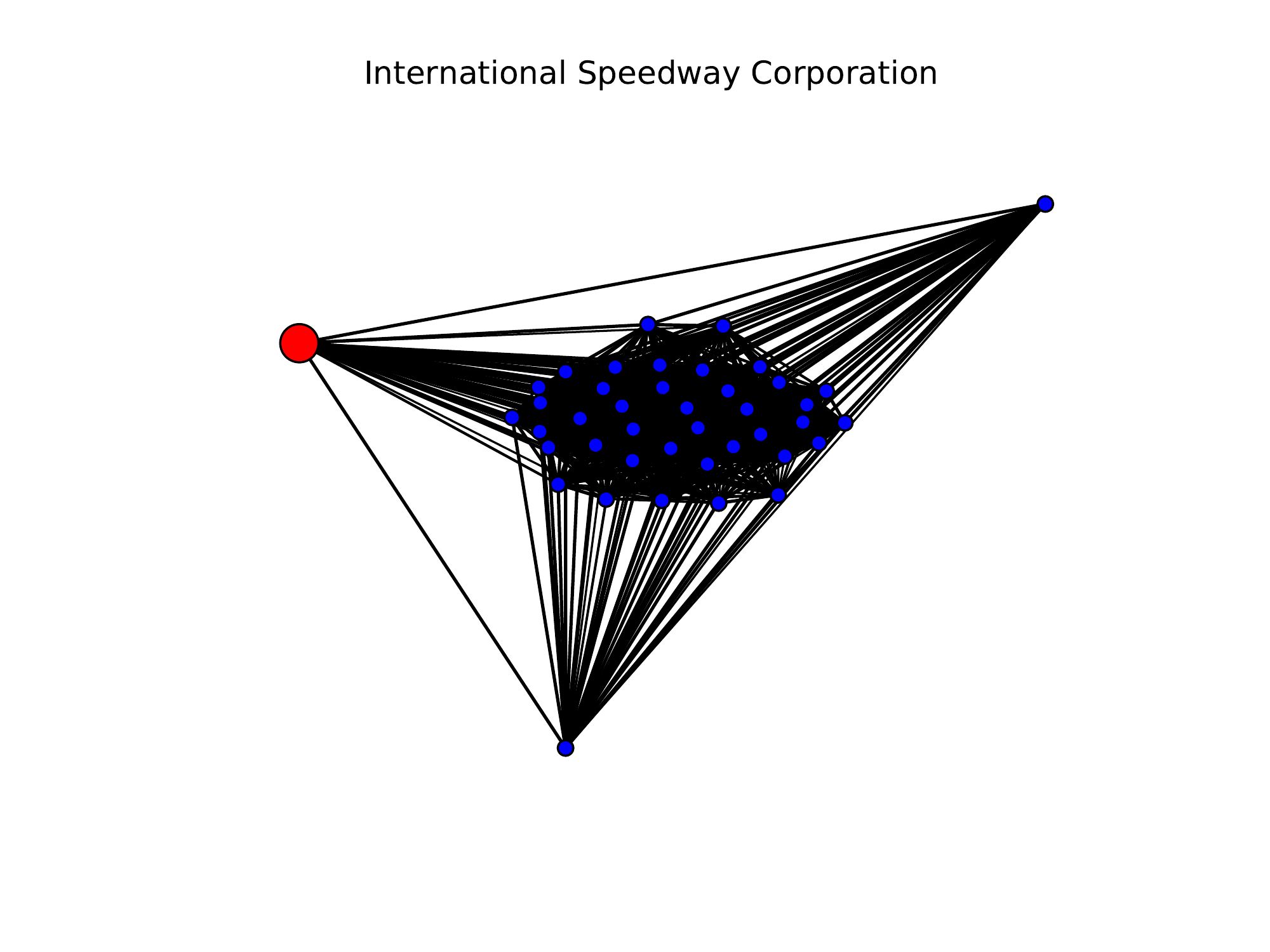}} 
 \subfigure[] {\includegraphics[width=70mm,height=50mm]{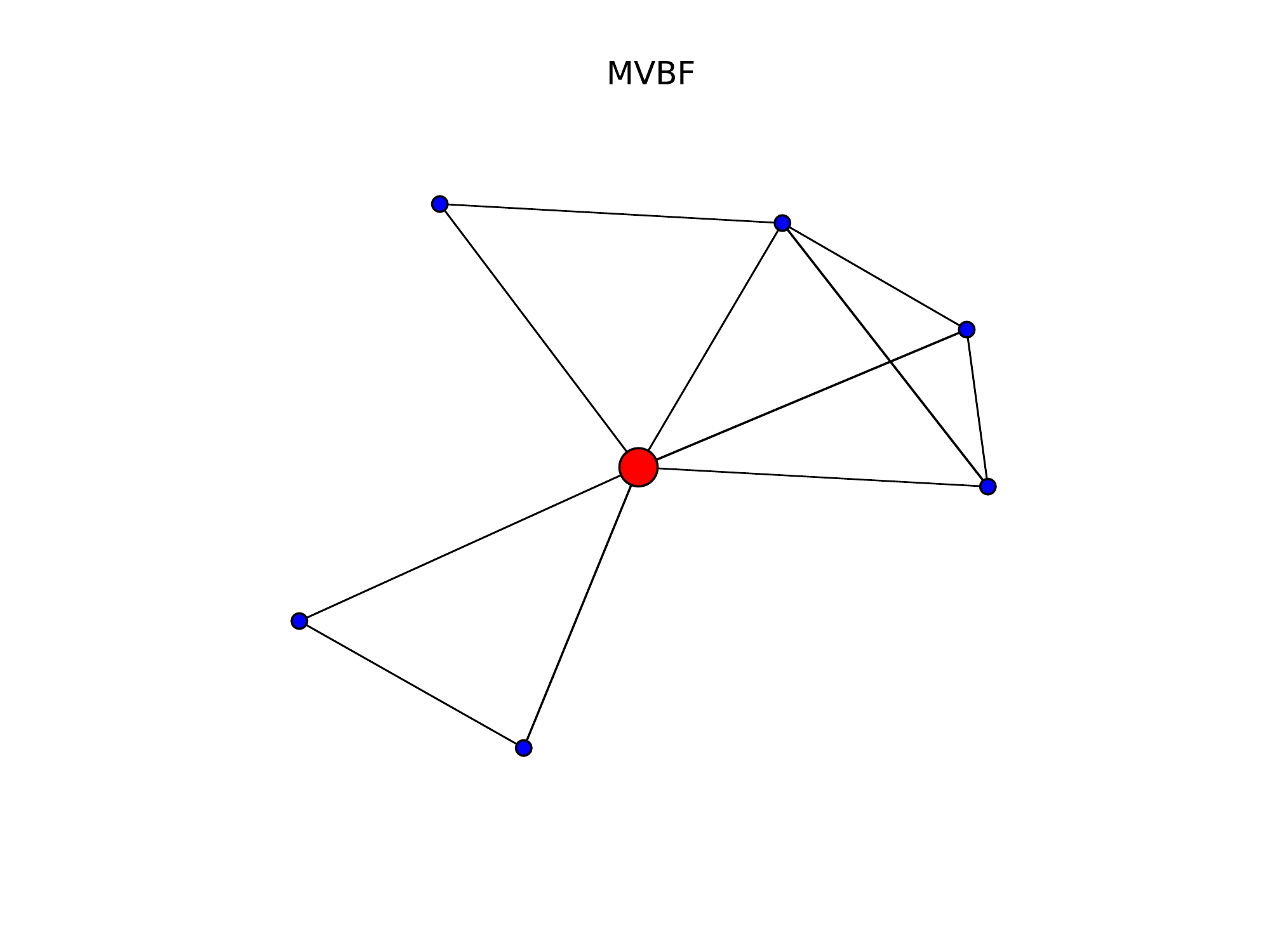}} 
\subfigure[]	 {\includegraphics[width=70mm,height=50mm]{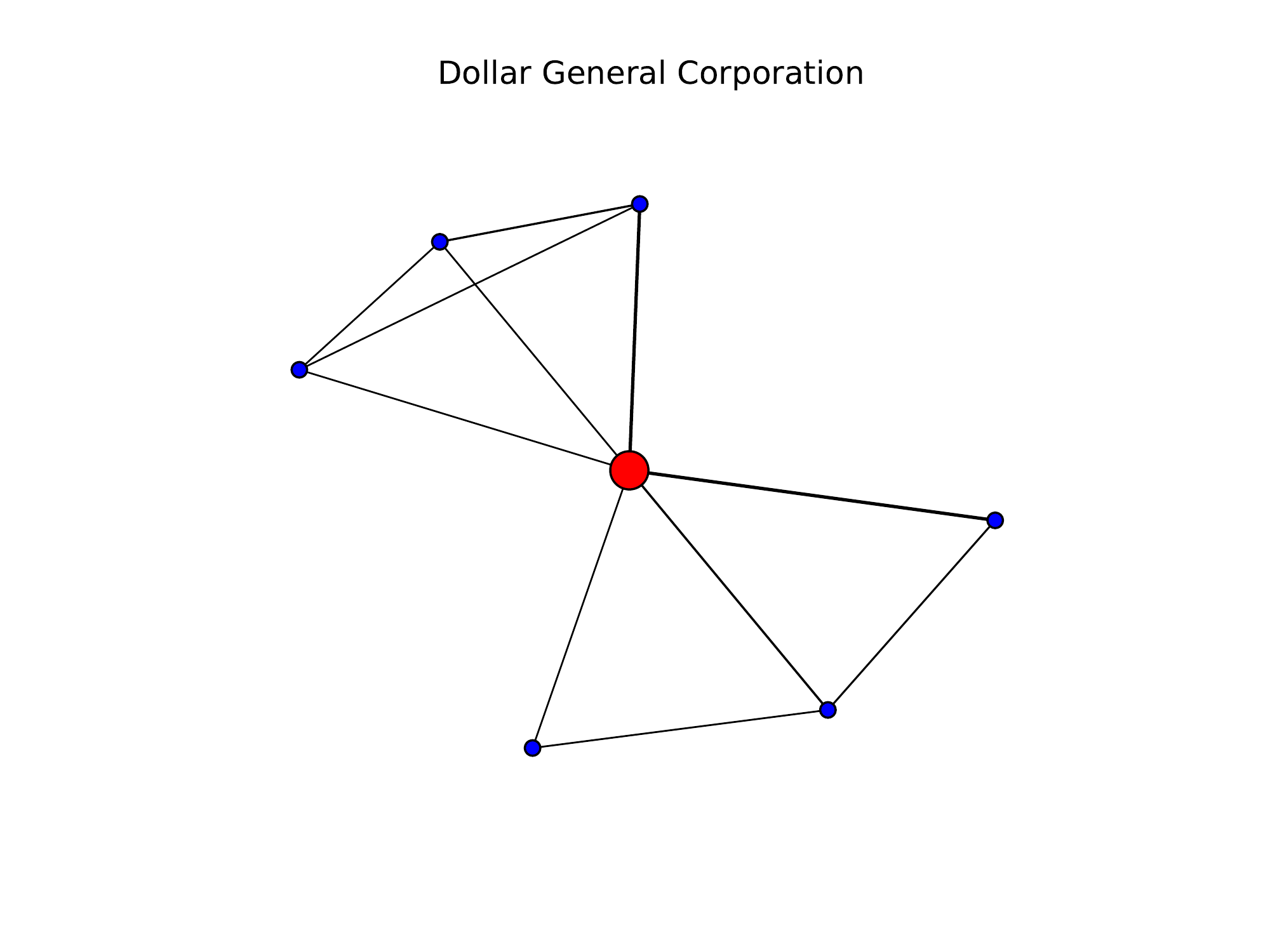} }
\subfigure[]	{\includegraphics[width=70mm,height=50mm]{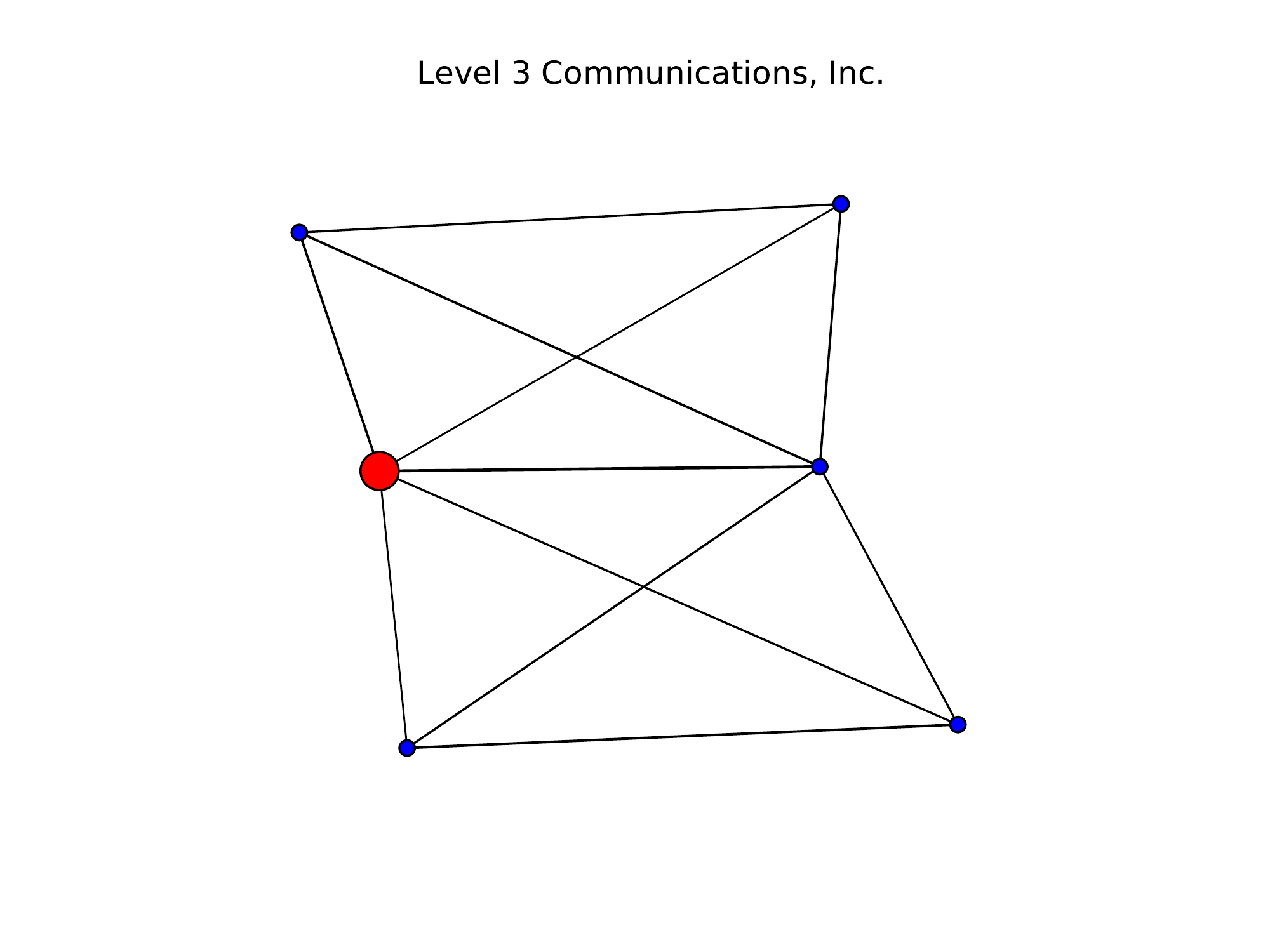}} \\
\end{center}
\caption{Egonets with Highest Outlier Scores: (a)-(b) Purchase; (c)-(d) Sale.}
\label{egonet}
\end{figure}

The least squares power law fitting, and corresponding top ten outliers are depicted in Figure~\ref{lsfit}.
The egonets of outlier ego nodes were identified and were found to follow an interesting pattern. The anomalous ego nodes often occupied a \textit{bridge} position between highly connected components, perhaps indicating the role of hubs between cliques (or quasi-cliques) of traders. 
 Examples of discovered anomalous egonets are shown in Figure~\ref{egonet}. 

\begin{figure}[h]
\begin{center}
\subfigure[] {  \includegraphics[width=70mm,height=55mm]{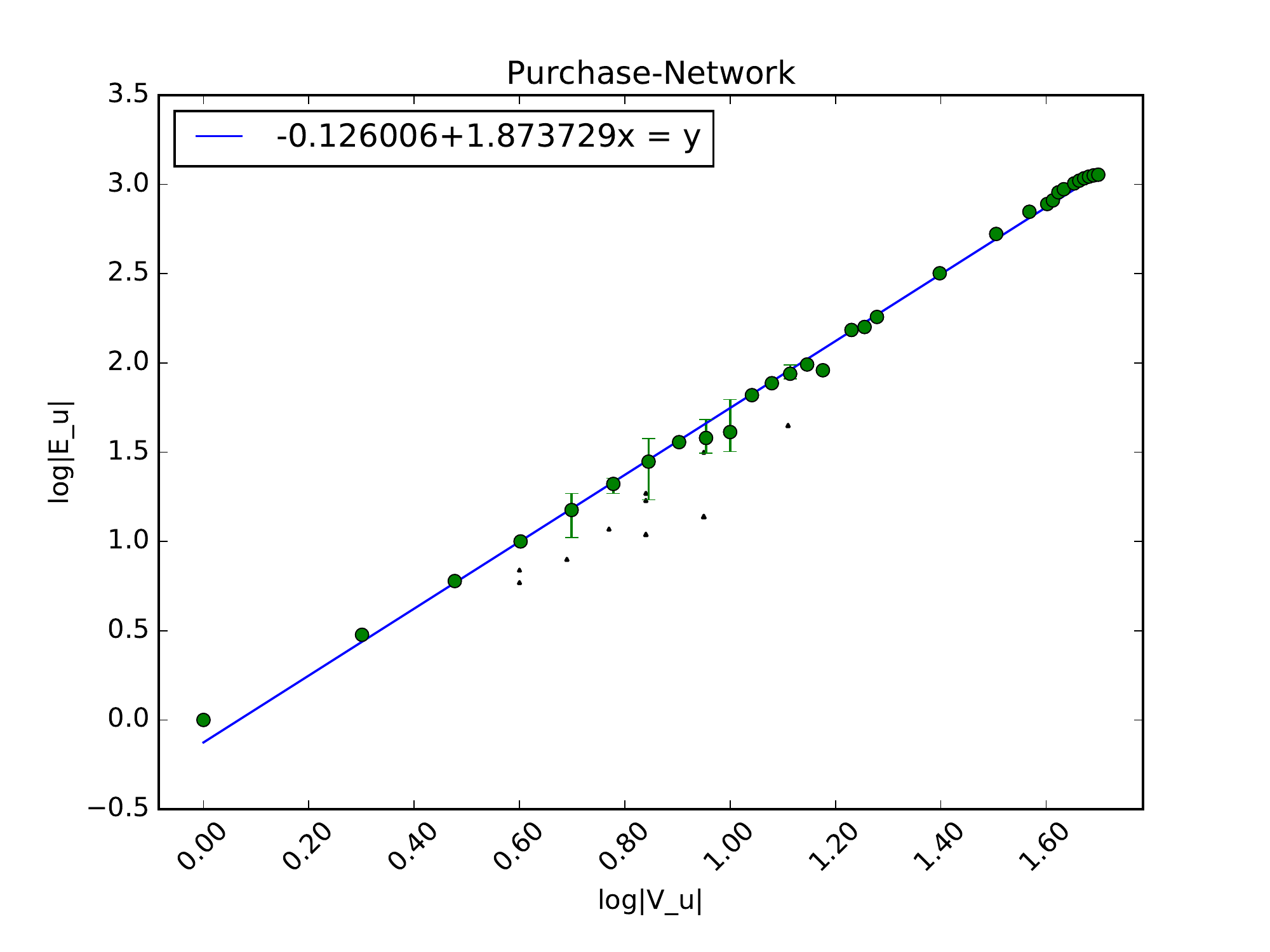}}
\subfigure[] {  \includegraphics[width=70mm,height=55mm]{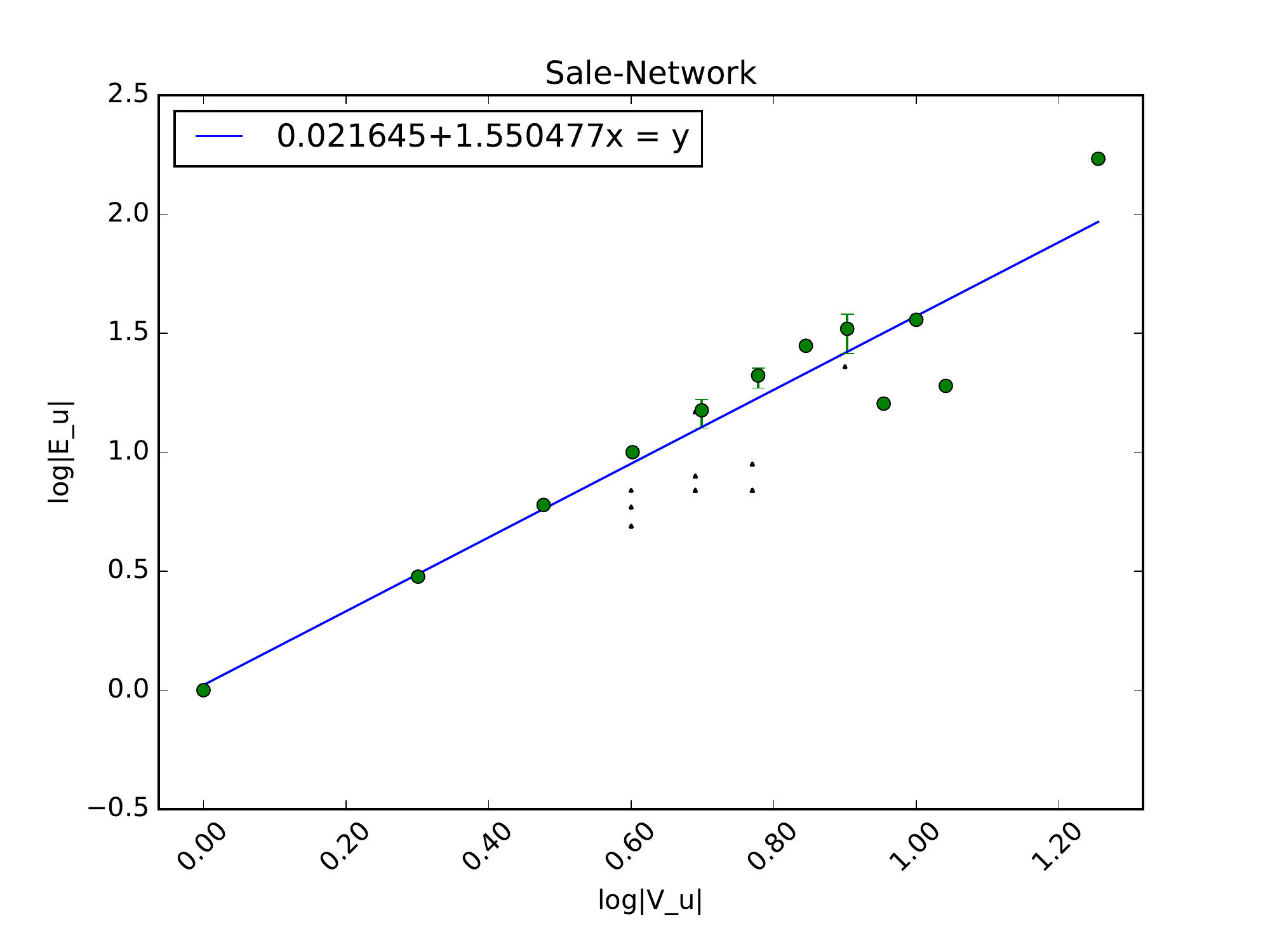}}
\end{center}
\caption{Power Law Fitting and Anomaly Detection (\textit{LCS-based}): (a) Purchase; (b)  Sale. }
\label{lcs_lsfit}
\end{figure}

\begin{figure}
\begin{center}
 \subfigure[] {\includegraphics[width=65mm,height=50mm]{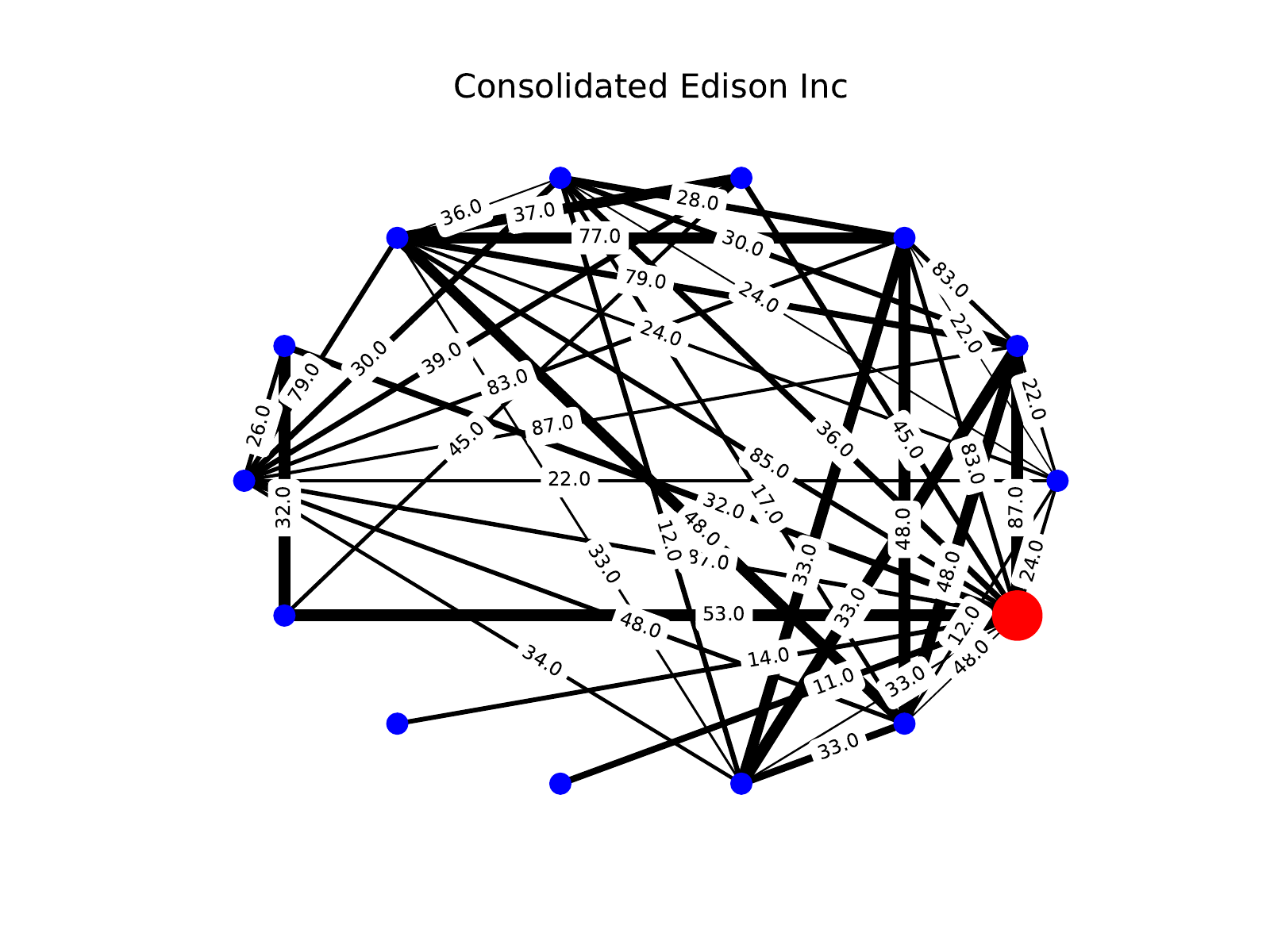}} 
 \subfigure[] {\includegraphics[width=70mm,height=50mm]{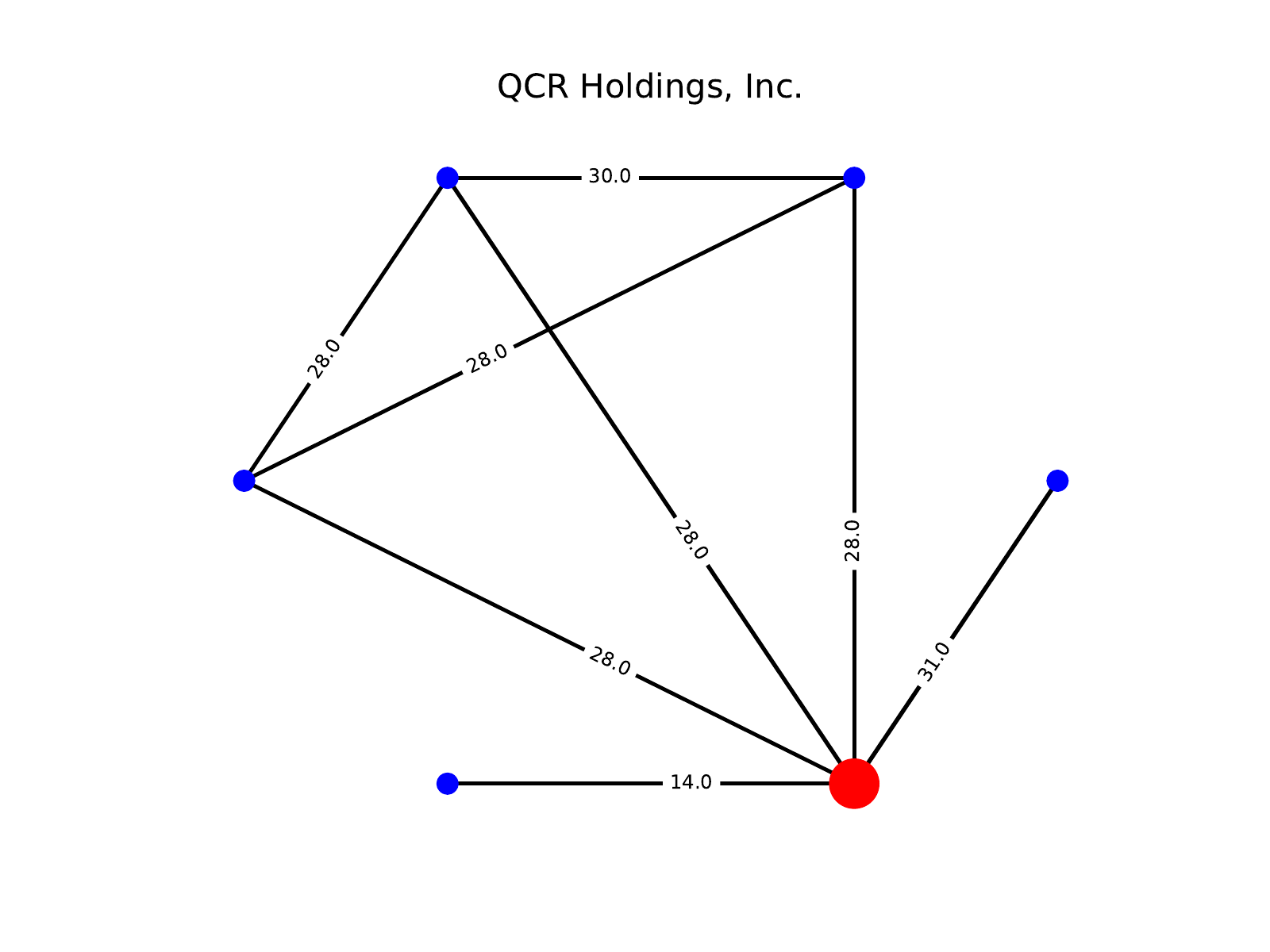}} 
\subfigure[]	 {\includegraphics[width=70mm,height=50mm]{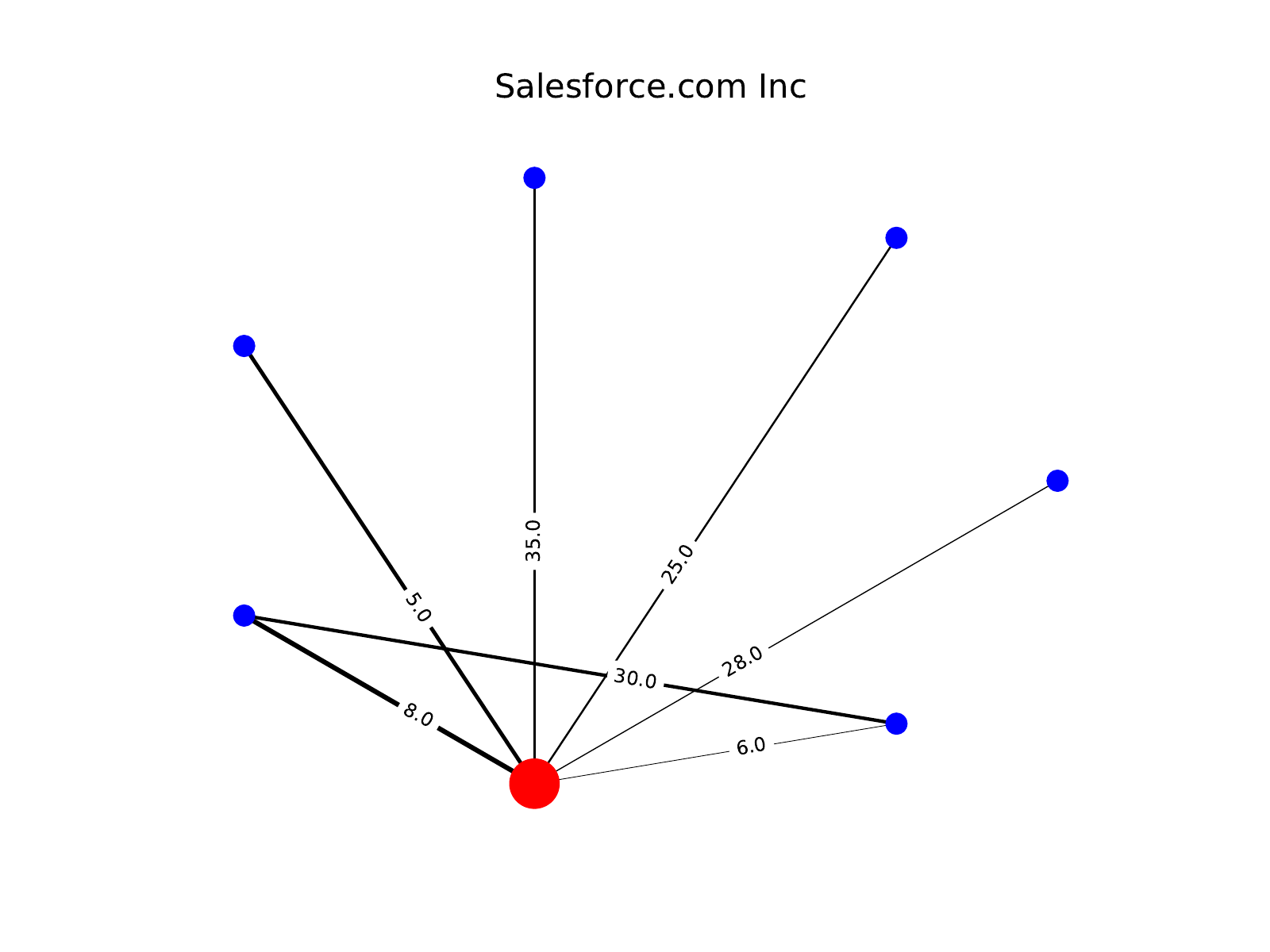} }
\subfigure[]	{\includegraphics[width=70mm,height=50mm]{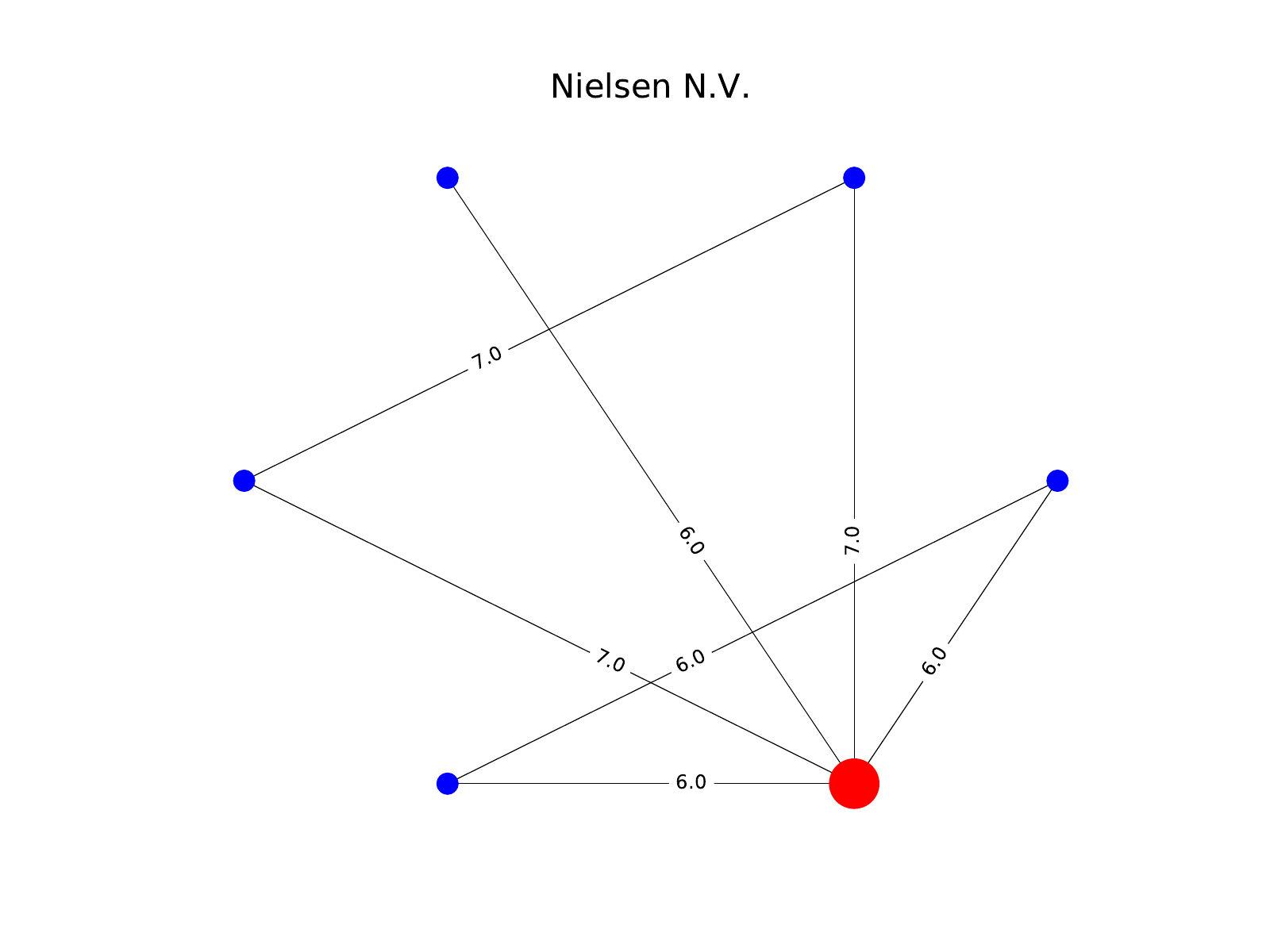}} \\
\end{center}
\caption{Egonets with Highest Outlier Scores (\textit{LCS-based}): (a)-(b) Purchase; (c)-(d) Sale. Edges are labeled with the \textit{LCS} length.}
\label{lcs_egonet}
\end{figure}

The least squares fitting and egonets for the \textit{LCS-based} network construction are shown in Figure~\ref{lcs_lsfit} and in Figure~\ref{lcs_egonet}. The \textit{LCS} length thresholds for the purchase and sale networks are set to 10 and 5 respectively. The edges of the egonets in Figure~\ref{lcs_egonet} are labeled with the length of the longest common sub-sequence shared by the corresponding two nodes.  The ego nodes detected by this method still largely manifest the role of hubs between cliques (or quasi-cliques).

\section{Hyper-graph-based Anomaly Detection}

\subsection{Motivation}  
\begin{figure}[h]
\begin{center}
\subfigure[] {  
\includegraphics[height=45mm]{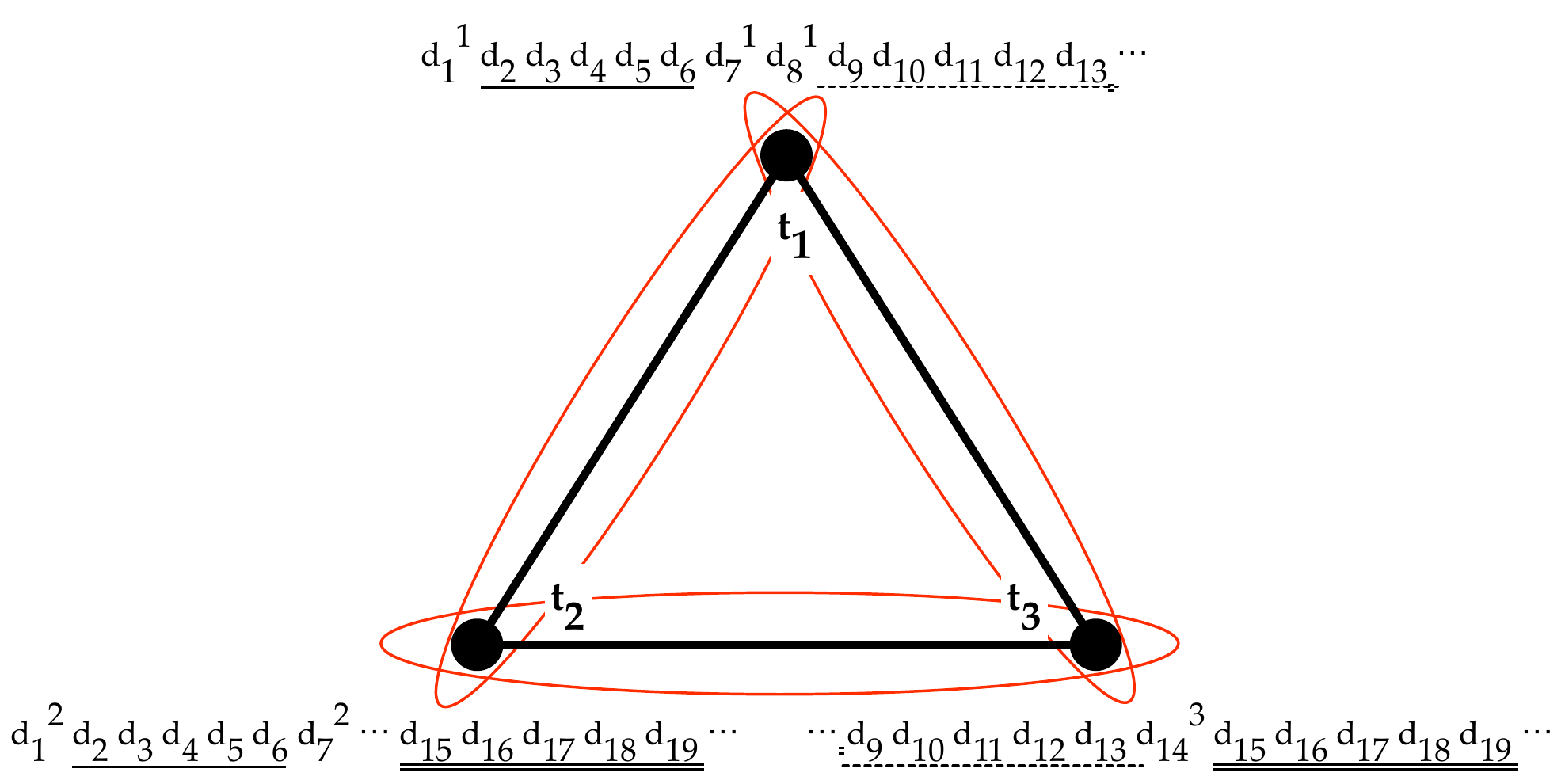}}\\
\subfigure[] {  \includegraphics[height=50mm]{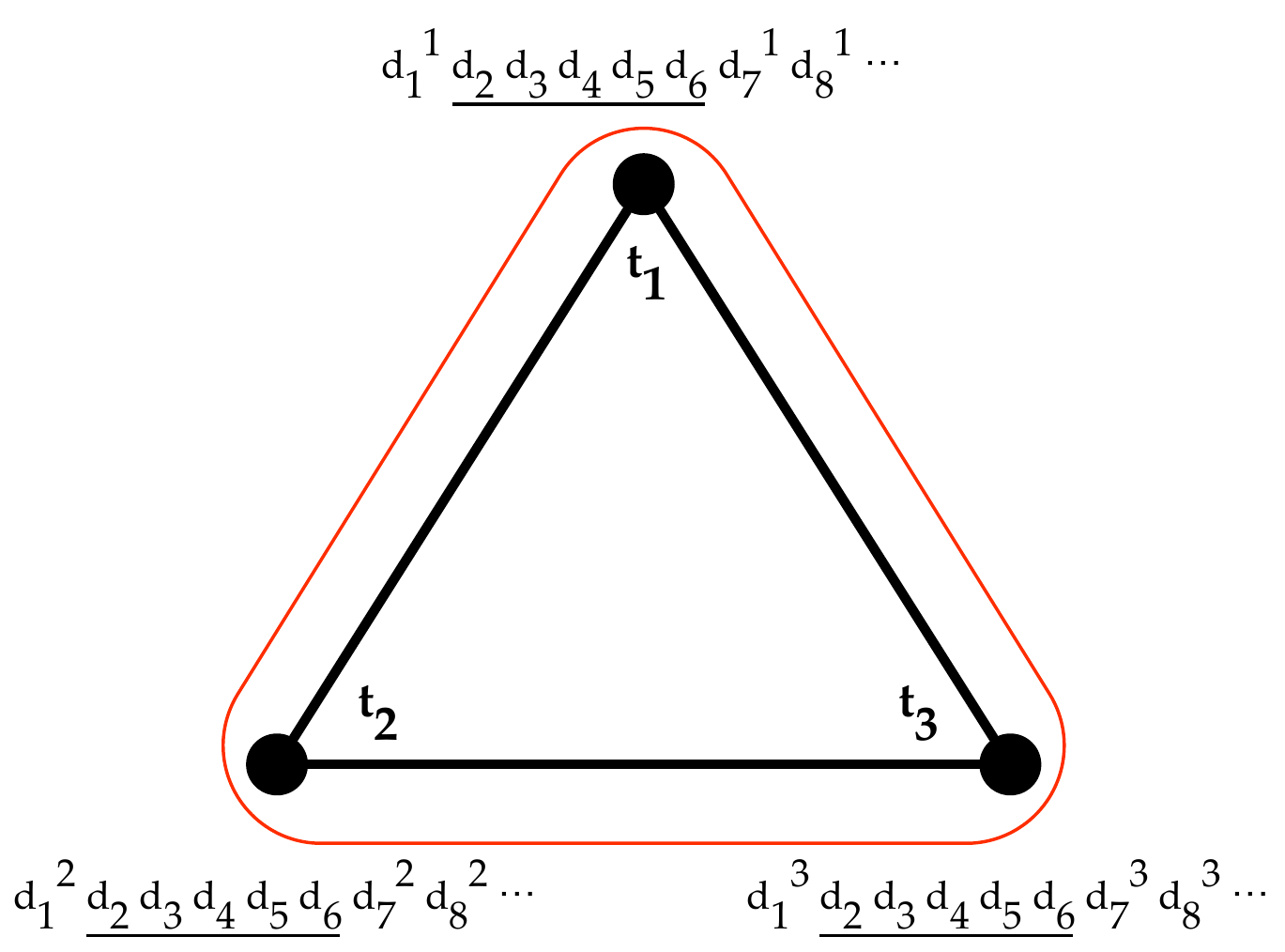}}
\end{center}
\caption{Graphs vs Hyper-graphs: (a) Three pairwise co-occurrences; (b) A three-way co-occurrence.}
\label{hyper-graph}
\end{figure}

While the graph-based methods discussed above lead to interesting results worth further investigation, they have a fundamental limitation rooted in the use of graphs to represent traders and their interactions. Graphs can only capture {\textit{pairwise}} interactions. As an example, consider the scenarios represented in Figure \ref{hyper-graph}. We have three traders $t_1$, $t_2$, and $t_3$. In (a), $t_1$ and $t_2$ share the sub-sequence of dates [$d_2$ $d_3$ $d_4$ $d_5$ $d_6$]; $t_1$ and $t_3$ share the sub-sequence [$d_9$ $d_{10}$ $d_{11}$ $d_{12}$ $d_{13}$]; and $t_2$ and $t_3$ share the sub-sequence [$d_{15}$ $d_{16}$ $d_{17}$ $d_{18}$ $d_{19}$]. Thus each pair of traders share a different sub-sequence of dates, resulting in a clique of size 3 (assuming $t=5$).
In (b), the three traders share the {\em{same}} sub-sequence of dates [$d_2$ $d_3$ $d_4$ $d_5$ $d_6$]. This results in the same 3-clique as in (a). Since a graph structure only captures pairwise co-occurrences, it is not able to distinguish the two scenarios, with a consequent loss of important information. In contrast, an hyper-graph can capture multi-way co-occurrences, and therefore can discriminate between the two scenarios. With hyper-graphs, case (a) is modeled with the three hyper-edges $\{t_1,t_2\}$, $\{t_1,t_3\}$, and $\{t_2,t_3\}$, while case (b) is modeled with the {\em{single}} hyper-edge $\{t_1,t_2,t_3\}$ (as depicted in red in Figure \ref{hyper-graph}). No loss of information is incurred this time.

\subsection{Building Hyper-graphs of Insiders}
To model multi-way interactions among traders, we constructed hyper-graphs from our data (purchase and sale). We used the LCS-based approach with the same thresholds as before. The resulting hyper-graph can be represented as $H=(V,E)$, where $V$ is a set of vertices (the traders) and $E$ is a set of hyper-edges, where each hyper-edge corresponds to a set of vertices.

\begin{figure}[h]
\centering
		\includegraphics[scale=0.4]{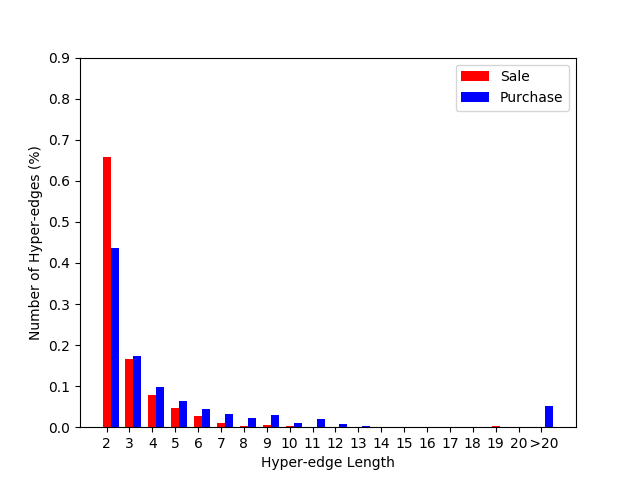}
		\caption{Distribution of Hyper-edges.}
		\label{hypergraph_stats}
		\end{figure}

\subsection{Preliminary Results} 
Figure~\ref{hypergraph_stats} shows the number of hyper-edges ($\%$) per size found in our data. Most of the hyper-edges are regular edges (i.e., size 2), and we observe an exponential decrease in number as the size increases, with a big gap between 2 and 3. This is an indication that hyper-edges capture an uncommon trading behavior which is worth exploring.  

\begin{figure}[h]
\begin{center}
\subfigure[] {  \includegraphics[width=65mm,height=40mm]{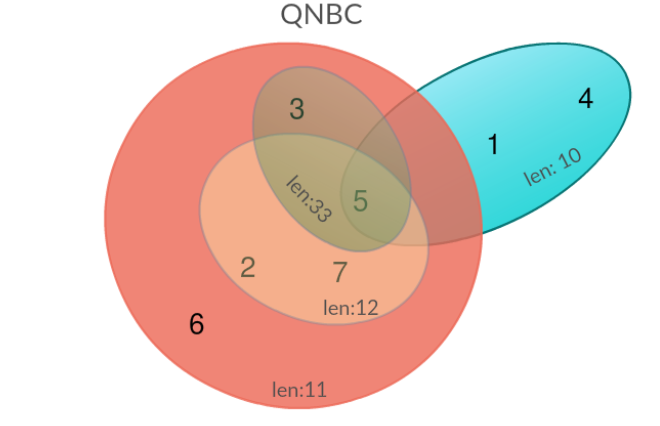}}
\subfigure[] {  \includegraphics[width=65mm,height=40mm]{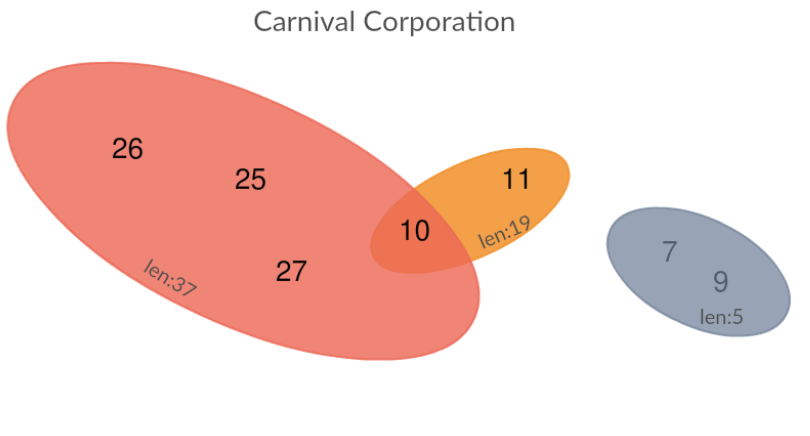}}
\end{center}
\caption{Examples of  Hyper-graphs: (a) Purchase; (b)  Sale. }
\label{hgraphs}
\end{figure}

Two examples of hyper-graphs obtained from our data are given in Figure~\ref{hgraphs}. The hyper-edges are annotated with the length of the corresponding \textit{LCS}.

Insiders that belong to the intersection of multiple hyper-edges, e.g. insider 5 in Figure~\ref{hgraphs} (a), correspond to hubs that share significant trading sequences with multiple cliques. We explore the characteristics of such insiders in the following to investigate their potential of being anomalies.  

\section{Evaluation}
A big challenge with the detection of illegal insider trading is evaluation. How do we quantify how informative the obtained results  are? How do we verify whether the identified traders have indeed operated illegally? Domain expertise and cross-checking against known past financial events (e.g., merging or splitting of corporations) are avenues to be explored to tackle this problem. 

In this work, to start evaluating our results we looked at the profit that the identified traders made during the {\textit{sequence of dates shared with the traders identified as similar}}. The intent was to see whether consistent profit was made on the transactions made during those days.  In particular, we computed the {\textit{signed normalized dollar amount}} as described in \cite{SNAM14}. Briefly, we compared the reported price of the transaction (purchase or sale) with the market closing price of the company's stock on the same day of the transaction. An insider makes a (positive) profit in two scenarios: when he buys shares of a stock at a price lower than the closing price for that stock, and similarly, when he sells shares of a stock at a price higher than the closing price. The amount is normalized by the dollar volume of the company's stock in question. Thus, the signed normalized dollar amount is a value between $-1$ and $1$. A positive value indicates a profit; a negative value indicates a loss.

\begin{figure}[h]
\begin{center}
\subfigure[] {  \includegraphics[width=60mm,height=40mm]{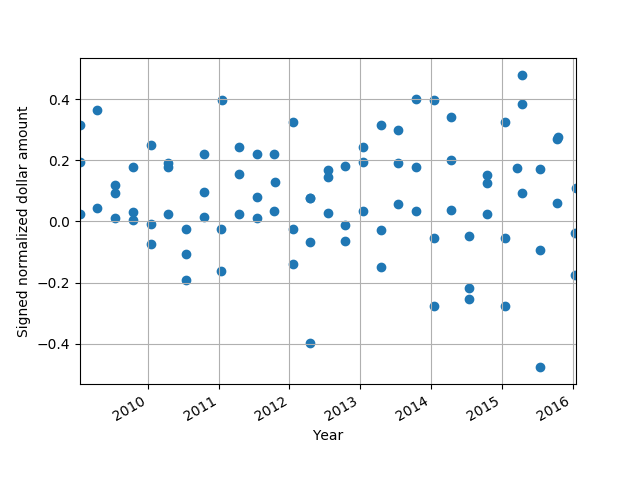}}
\subfigure[] {  \includegraphics[width=60mm,height=40mm]{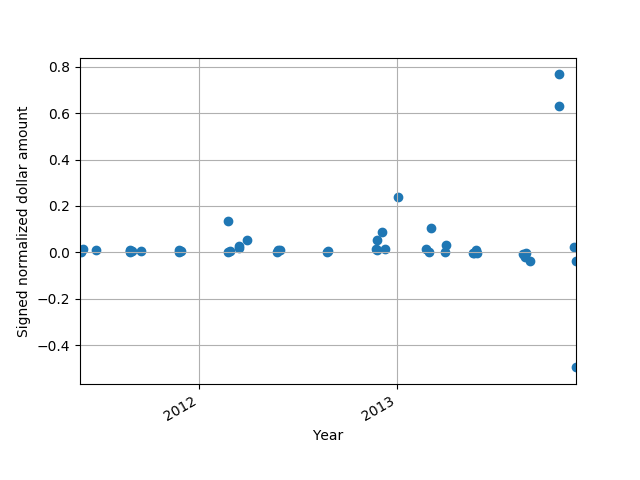}}
\end{center}
\caption{Time series of the signed normalized dollar amounts (LCS-based Egonets) : (a) Purchase; (b)  Sale.}
\label{dvplot_egonets}
\end{figure}

We first computed  the signed normalized dollar amounts for the top ranked insiders identified using the LCS-based ego nets. All the hyper-edges which included the insider node were computed, and the union of the corresponding trading sequences were used to compute the time series of the normalized dollar amounts. Results  are given in Figure~\ref{dvplot_egonets}. Unfortunately, trading prices were not always available in the data. As such, those reported are a subset of the total actual transactions. We observe that in both cases (purchase and sale), the majority of the transactions are located above the 0 level, which is an indication of repeated profit. Furthermore, Figure~\ref{dvplot_egonets} (b) shows two transactions that resulted in a very large profit.

\begin{figure}[h]
\begin{center}
\subfigure[] {  \includegraphics[width=60mm,height=40mm]{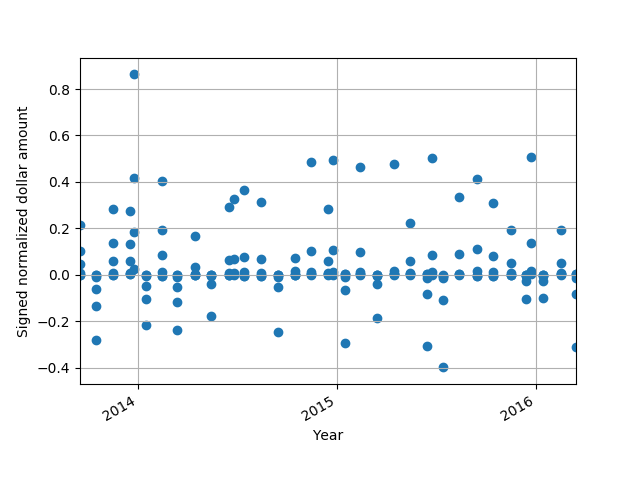}}
\subfigure[] {  \includegraphics[width=60mm,height=40mm]{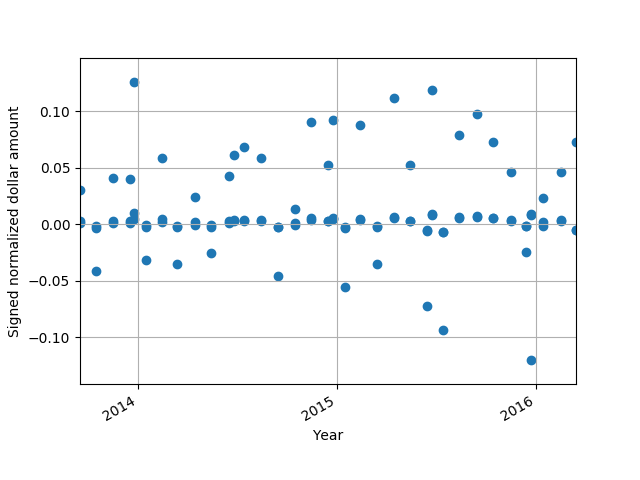}}
\subfigure[] {  \includegraphics[width=60mm,height=40mm]{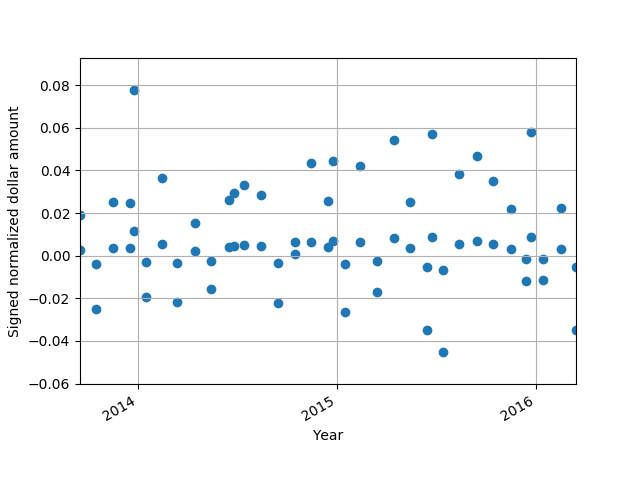}}
\subfigure[] {  \includegraphics[width=60mm,height=40mm]{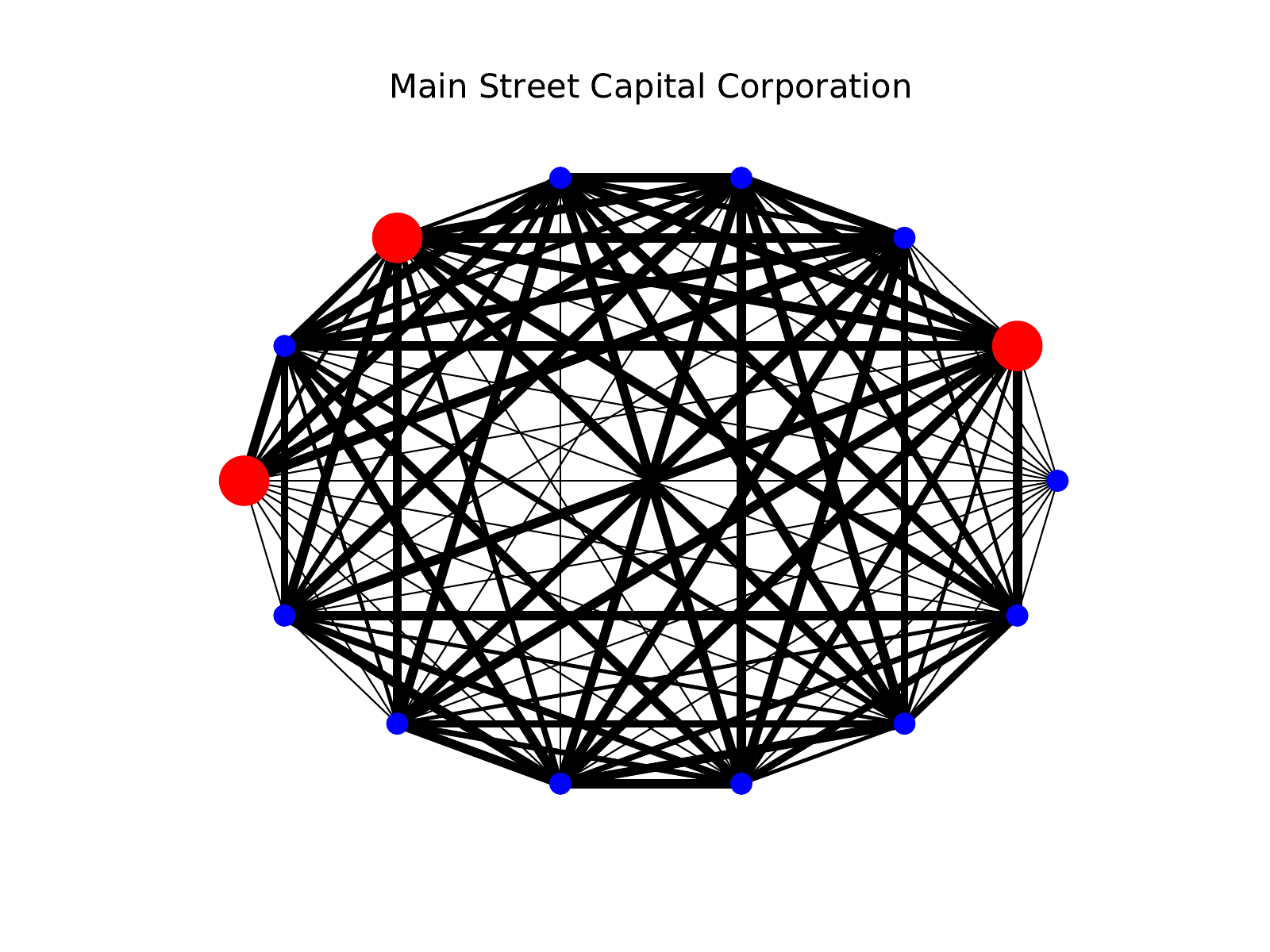}}
\end{center}
\caption{(a)-(c) Time series of the signed normalized dollar amounts of three insiders sharing multiple hyper-edges (Hyper-graph based); (d) Egonet which includes the three insiders. The thickness of the edges is proportional to the length of the shared sub-sequence.}
\label{dvplot_hgraph}
\end{figure}

As discussed in the previous section, we also plotted the time series of the signed normalized dollar amounts for insiders of the hyper-graphs which lie at the intersection of multiple hyper-edges (at least 4 in our experiments). Figure~\ref{dvplot_hgraph} shows a sample result of the time series for a group of three insiders at the intersection of multiple hyper-edges and belonging to the same egonet (also shown). Again, the majority of the transactions is above the 0 level.

\section{Conclusion and Future Work}

In this work, we have collected and analyzed insider trading data. To capture the relationship between trading behaviors of insiders, we have constructed different kinds of graphs. 

Our results suggest the discovery of interesting
patterns. The anomalous ego nodes we have identified often occupied a bridge position between highly connected components, perhaps indicating the role of hubs between cliques of traders. Our anomaly ranking can be used by investigators to prioritize cases for further analysis. We have also argued for the need of higher-order structures, and therefore captured multi-way interactions among insiders via the construction of hyper-graphs. The relevance of the identified cases is supported  by the analysis of the dollar amount time series signifying profit. As discussed earlier, more work is needed to develop a thorough evaluation methodology.

We believe the complex patterns captured by hyper-graphs of insiders deserve further exploration. We are considering a model-based generative approach to learn the distributions underlying normal vs. anomalous hyper-edges. Potential meta-features to be considered are the size of the hyper-edges and the characterizing sequences of dates (including their length). Parameters can be estimated with a variational EM approach. A similar method was introduced in \cite{Silva09} in a different context, but the anomalous distribution is  assumed to be known and fixed, and no meta-features are taken into consideration.



\begin{thebibliography}{99}



\bibitem{SNAM14}
A. Tamersoy, E. Khalil, B. Xie, S.~L. Lenkey, B.~R. Routledge, D.~H. Chau, and S.~B. Navathe, 
{\em Large Scale Insider Trading Analysis: Patterns and Discoveries}, 
Social Network Analysis and Mining (SNAM), 4(1), 1-17 (2014).

\bibitem{Goldberg03}
H.~G. Goldberg, J.~D. Kirkland, D. Lee, P. Shyr, and D. Thakker, 
{\em The NASD securities observation, new analysis and regulation system (SONAR)},
In Proceedings of the Conference on Innovative Applications of Artificial Intelligence (2003).

\bibitem{Donoho}
S. Donoho,  
{\em Early detection of insider trading in option markets}, In Proceedings of the ACM SIGKDD Conference on Knowledge Discovery and Data Mining (2004).

\bibitem{Akoglu2010}
L. Akoglu, 	M. McGlohon, and C. Faloutsos,
{\em OddBall: spotting anomalies in weighted graphs},
In Proceedings of the 14th Pacific-Asia conference on Advances in Knowledge Discovery and Data Mining - Volume Part II, Pages 410-421 (2010).

\bibitem{LOF}
M.~M. Breunig, H.~P. Kriegel, R.~T. Ng, and J. Sander,
{\em LOF: Identifying Density-Based Local Outliers},
In Proceedingd of the International Conference on Management of Data, (2000).

\bibitem{Akoglu-survey}
L. Akoglu, 	H. Tong, and D. Koutra,
{\em Graph based anomaly detection and description: a survey}, Data Mining and Knowledge Discovery, Volume 29, Issue 3, Pages 626-688, (2015).

\bibitem{Silva09}
J. Silva, and R. Willett, 
{\em Hypergraph-Based Anomaly Detection of High-Dimensional Co-Occurrences},
IEEE Transactions on Pattern Analysis and Machine Intelligence,vol. 31, no. 3, (2009).

%
%
%
%
%
%
%
%
%

\end{thebibliography}
\end{document}